\documentstyle[preprint,aps]{revtex}
\def\to{\rightarrow}
\begin{document}
\draft
\preprint{\vbox{\hbox{IFT--P.008/96}\hbox{hep-ph/9603343}}}
\title{Anomalous Higgs Boson Contribution to $e^+ e^- \to b \bar{b} 
\gamma$ at LEP2. }
\author{ S.\ M.\ Lietti, S.\ F.\ Novaes and R.\ Rosenfeld}
\address{Instituto de F\'{\i}sica Te\'orica, 
Universidade  Estadual Paulista, \\  
Rua Pamplona 145, CEP 01405-900 S\~ao Paulo, Brazil.}
\date{\today}
\maketitle
\widetext
\begin{abstract}
We study the effect of anomalous $H\gamma\gamma$ and $HZ\gamma$
couplings,  described by a general effective Lagrangian, on the
process  $e^+ e^- \to b \bar{b} \gamma $ at LEP2 energies. We
include the relevant irreducible standard model background to
this process, and from the photon energy spectrum, we determine
the reach of LEP2 to unravel the  anomalous couplings by
analyzing the significance of the signal for Higgs boson with
mass up to $150$ GeV.
\end{abstract}

\section{Introduction}
\label{sec:int}

The Standard Model (SM) has been tested to an unprecedent degree
of accuracy of $0.1 \%$ in some of the physical observables at
LEP1, with many implications to physics beyond the SM
\cite{Hagiwara}. However, the $Z$--pole experiments are able to
probe with great precision just the fermionic couplings to the
vector bosons while furnishing very little information about the
interaction between the gauge bosons and the Higgs sector of the
SM. In principle, it is conceivable that the interactions of the
Higgs boson, which is responsible for the spontaneous breaking of
the electroweak symmetry and for generating fermion masses, are
different from those prescribed by the SM. In this case, an
effective Lagrangian formalism can be used to describe possible
anomalous interactions between the Higgs boson and the vector
bosons.

The effective Lagrangian approach is a convenient model--indepent
parametrization of the low--energy effects of new physics beyond
the SM that may show up at higher energies \cite{effective}.
Effective Lagrangians, employed to study processes at a typical
energy scale $E$, can be written as a power series in
$1/\Lambda$, where the scale $\Lambda$ is associated with the new
particle masses belonging to the underlying theory.  The
coefficients of the different terms in the effective Lagrangian
arise from integrating out the heavy degrees of freedom that are
characteristic of a particular model for new physics. Invariant 
amplitudes, generated by such Lagrangians, will be an expansion in
$E/\Lambda$, and in practice one can only consider the
first few terms of the effective Lagrangian, {\it e.g.} dimension
six operators, which are dominant for $E \ll \Lambda$.

The anomalous $H\gamma\gamma$ and $HZ\gamma$ couplings have
already been  considered in $Z$ and Higgs decays
\cite{Hagiwara2}, in $e^+ e^-$ collisions \cite{Hagiwara2,epem}
and at $\gamma\gamma$ colliders \cite{gg}. In this paper, we
concentrate our analyses on the effect of anomalous
$H\gamma\gamma$ and $HZ\gamma$ couplings, described by a general
effective Lagrangian, on the process  $e^+ e^- \to b \bar{b}
\gamma $. This is a very interesting reaction since the SM
contribution to $e^+ e^- \to H ( \to b \bar{b}) \gamma$, being a
one loop process,  is extremelly small, and the observation of
any $H\gamma$ event at LEP2 will be a clear signal of new
physics. It may also be the only possibility of detecting a Higgs
boson with mass larger than $80$ GeV at LEP2, provided that the
anomalous couplings are sufficiently large. In our calculation,
we include the dominant decay of the Higgs boson into a pair of
bottom and anti--bottom quarks in the framework of a complete
tree--level calculation of the process $e^+ e^- \to b \bar{b}
\gamma$ involving both the SM and  the anomalous Higgs boson
couplings. In this way, the  irreducible SM background to the new
physics is taken into account. Furthermore, we employ the photon
energy spectrum to identify the existence of the anomalous Higgs
boson or at least impose further bounds on its effective
couplings.

This paper is organized as follows: in Sec.\ \ref{sec:eff}, we
review the use of effective Lagrangians to study anomalous Higgs
boson couplings, including limits on these couplings arising from
precision measurements. In Sec.\ \ref{sec:pro},  we study the
process $e^+ e^- \to b \bar{b} \gamma$, and we present our
results and trace our conclusions in Sec.\ \ref{sec:res:con}.

\section{Effective Lagrangians and the anomalous  $H\gamma\gamma$ and
$HZ\gamma$ couplings} 
\label{sec:eff}

In order to define an effective Lagrangian, it is necessary to
specify the symmetry and the particle content of the low--energy
theory. In our case, we require the effective Lagrangian to be
CP--conserving, invariant under the  SM symmetry $SU(2)_L \times
U(1)_Y$ and to have as fundamental fields the same ones 
appearing in the SM spectrum. In particular, the Higgs field will be
manifest and the symmetry is realized linearly. There are eleven
independent dimension-six operators \cite{dim6} of which only
five are relevant for our discussions. Following the notation of
reference \cite{dim6:zep}, we can write, 
\begin{eqnarray}
{\cal L}_{eff} = \frac{1}{\Lambda^2} &&\left[  
            f_{BW} \Phi^{\dagger} \hat{B}_{\mu \nu} \hat{W}^{\mu \nu} \Phi +
f_W (D_{\mu} \Phi)^{\dagger} \hat{W}^{\mu \nu}  (D_{\nu} \Phi) +
f_B  (D_{\mu} \Phi)^{\dagger} \hat{B}^{\mu \nu}  (D_{\nu} \Phi)  \right. 
\nonumber \\
 & & \left. \; + f_{WW} \Phi^{\dagger} \hat{W}_{\mu \nu} 
\hat{W}^{\mu \nu} \Phi +
f_{BB}  \Phi^{\dagger} \hat{B}_{\mu \nu} \hat{B}^{\mu \nu} \Phi
\right]
\; , 
\label{eff}
\end{eqnarray}
where $\Phi$ is the Higgs field doublet, which in the unitary
gauge assumes the form,
\[
\Phi = \left(\begin{array}{c}
0 \\
(v + H)/\sqrt{2}
\end{array}
\right) \; , 
\]
and 
\begin{equation}
 \hat{B}_{\mu \nu} = i \frac{g'}{2} B_{\mu \nu}  \;\; , \;\; \; 
\hat{W}_{\mu \nu} = i \frac{g}{2} \sigma^a W^a_{\mu \nu} \; , 
\label{lagr}
\end{equation}
with $B_{\mu \nu}$ and $ W^a_{\mu \nu}$ being the field strength
tensors of the respective $U(1)$ and $SU(2)$ gauge fields.

This Lagrangian gives rise to the following anomalous
$H\gamma\gamma$ and $HZ\gamma$ couplings, in the unitary gauge
\cite{Hagiwara2},
\begin{equation}
{\cal L}_{eff}^{H \gamma \gamma, H Z \gamma} = 
 g_{H \gamma \gamma}
H A_{\mu \nu} A^{\mu \nu} + g^{(1)}_{H Z \gamma} A_{\mu \nu} Z^{\mu} 
\partial^{\nu} H + g^{(2)}_{H Z \gamma} H A_{\mu \nu} Z^{\mu \nu}
\; , 
\label{H} 
\end{equation} 
where $A(Z)_{\mu \nu} = \partial_\mu A(Z)_\nu - \partial_\nu
A(Z)_\mu$, and  the coupling constants $g_{H \gamma \gamma}$, and
$g^{(1,2)}_{H Z \gamma}$ are related to the coefficients of the
operators appearing in (\ref{lagr}) through, 
\begin{eqnarray}
g_{H \gamma \gamma} &=& - \left( \frac{g M_W}{\Lambda^2} \right)
                       \frac{s^2 (f_{BB} + f_{WW} - f_{BW})}{2} \; , 
\nonumber \\
g^{(1)}_{H Z \gamma} &=& \left( \frac{g M_W}{\Lambda^2} \right) 
                     \frac{s (f_W - f_B) }{2 c} \; ,  \\
\label{g}
g^{(2)}_{H Z \gamma} &=& \left( \frac{g M_W}{\Lambda^2} \right) 
                      \frac{s (2 s^2 f_{BB} - 2 c^2 f_{WW} + 
                     (c^2-s^2)f_{BW} ) }{2 c} \nonumber \; , 
\end{eqnarray}
with $g$ being the electroweak coupling constant, and $s(c)
\equiv \sin(\cos)\theta_W$.

The coefficients $f_B$ and $f_W$ can be related to triple vector
boson anomalous couplings and are bounded, for instance, by the
direct measurement of $WW\gamma$ vertex at hadron colliders.
However, more stringent bounds on the coefficients of the
effective Lagrangian (\ref{eff}) come from the precision
measurements of the electroweak parameters obtained at LEP1
\cite{dim6:zep}. Typically one has that $|f_{W, B, WW,
BB}/\Lambda^2|$ can be as large as $100$ TeV$^{-2}$, whereas
$|f_{BW}/\Lambda^2|$ should be at most $\sim 1$ TeV$^{-2}$.

\section{Anomalous Couplings and Higgs Boson Contribution to
$\lowercase{e}^+ \lowercase{e}^- \to \lowercase{b}
\bar{\lowercase{b}} \gamma$}
\label{sec:pro}

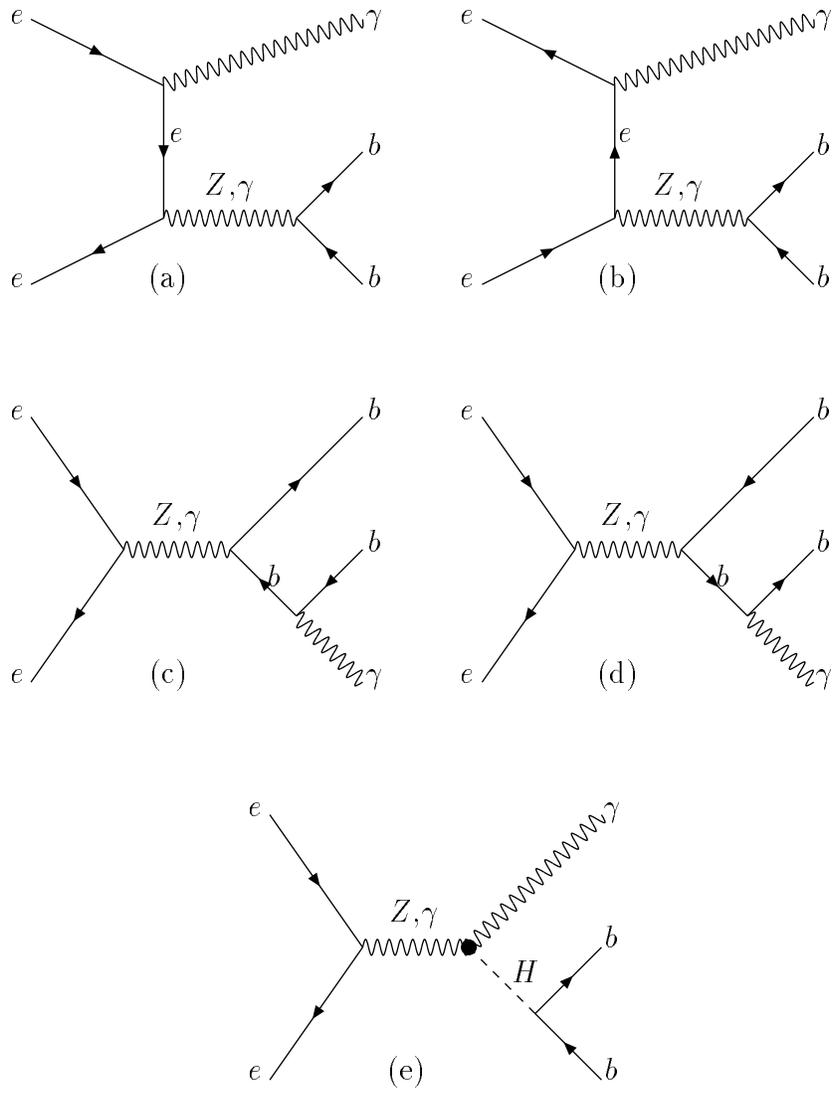
\begin{figure}
\protect
\caption{Feynman diagrams for  $e^+ e^- \to b \bar{b} \gamma$ in
the standard model at tree--level (a, b, c, d) and the anomalous Higgs 
boson contribution (e).}
\label{fig:1}
\end{figure}

An interesting option to test the couplings described by
(\ref{H}) at LEP2 is via the reaction $e^+ e^- \to H \gamma$,
with the subsequent decay of the Higgs boson into a $b
\bar{b}$ pair. In the SM, at tree-level, there are eight Feynman
diagrams that contribute to the process $e^+ e^- \to b \bar{b}
\gamma$ (see Fig.\ \ref{fig:1}$(a)$ -- $(d)$). 

A  SM Higgs boson contribution to this process appears only at
one--loop level, and is extremely small. For instance, the total
cross setion for the process  $e^+ e^- \to H \gamma$ \cite{H:g},
at $\sqrt{s} \simeq 175$ GeV, varies from $0.2$ fb to $0.02$ fb, for
the Higgs mass in the range $70 < M_H < 150$ GeV. Therefore, with
the expected LEP2 luminosity, no such events should be seen. Even
in the Minimal Supersymmetric Standard Model, one cannot expect
an enhancement larger than a factor of $3$ with respect to the SM
result  \cite{foot}. In this way, we neglect this loop
contribution in our calculation.

The bulk of the SM cross section comes from the $Z$ boson
contribution to the diagrams $(a)$ -- $(b)$ when the $Z$ boson is
on-mass-shell, and the process is effectively a $2$-body one.
This implies that  the majority of the photons emitted are
monochromatic, with energy given by $E_{\gamma}^Z = (s -
M_Z^2)/(2 \sqrt{s})$. 

When we take into account the anomalous Higgs boson couplings
described above, two additional diagrams should be considered
(see Fig.\ \ref{fig:1}(e)). Their contributions are dominated by
the on--mass--shell $H \gamma$ production, with $H \to b
\bar{b}$. Therefore, we can anticipate the existence of a a
secondary peak in the photon energy spectrum, generated at an
energy 
\begin{equation}
E_{\gamma}^H = \frac{s - M_H^2}{2 \sqrt{s} }  
\label{ehiggs}
\end{equation}
which would be a very clear signal for the Higgs boson.

In order to evaluate the total cross section and kinematical
distributions for the process  $e^+ e^- \to b \bar{b} \gamma$, we
have used the package MadGraph \cite{Madgraph} coupled to DHELAS,
the double precision version of HELAS \cite{helas},  for
generating the tree--level SM amplitudes.  We have written the
relevant subroutines for the Higgs anomalous couplings, and
included in the MadGraph generated file the two additional
anomalous amplitudes. In this way, all interference effects
between the SM and the anomalous  amplitudes were taken into
account. We checked for electromagnetic gauge invariance of the
whole invariant amplitude, and incorporated a three--body phase
space code,  based on \cite{collider}. Since the Higgs boson
resonance is very narrow, $\Gamma ( H \to b \bar{b} ) \sim 5$
MeV, for $M_H \sim 100$ GeV, we make sure to use appropriate
variables to take care of the Higgs events close to the resonance
peak. Finally,  we used VEGAS \cite{vegas} to perform the phase
space integration.

In our analyses we have assumed a center--of--mass energy of
$\sqrt{s} = 175$ GeV for the LEP2 collider, with a luminosity of
$0.5$ fb$^{-1}$. Our results were obtained using the
following energy and angular cuts, 
\begin{eqnarray}
E_{\gamma} &\geq& 20 \; \mbox{GeV} \; , \label{emin} \\
| \cos \theta_{e^- (e^+) \gamma } | & \leq & 0.87 \; , \\
| \cos \theta_{b (\bar{b}) \gamma } | & \leq & 0.94 \; .
\end{eqnarray} 
The photon energy cut (\ref{emin}) is intended to reject the 
background from  unresolved pair of photons from $\pi^0$ decays 
and assures, in principle, a  sensitivity to $M_H$ up to $150$ GeV.
The cuts in  $\cos \theta_{e^- (e^+) \gamma }$ and $\cos
\theta_{b (\bar{b}) \gamma }$ were introduced to reduce initial
and final state radiation, respectively. 

\section{Results and Conclusions}
\label{sec:res:con}

Our purpose is to determine the range of anomalous
$H\gamma\gamma$ and $HZ\gamma$  couplings that could be probed at
LEP2 by searching for a signal of the Higgs boson in the process
$e^+ e^- \to b \bar{b} \gamma$. We assume that the Higgs
couplings to fermions are the standard ones, which makes the
$BR(H\to  b \bar{b})$ dominant in the range $70 < M_H < 150$ GeV,
for $|f_i/\Lambda^2| \sim $ TeV$^{-2}$ \cite{Hagiwara2}.

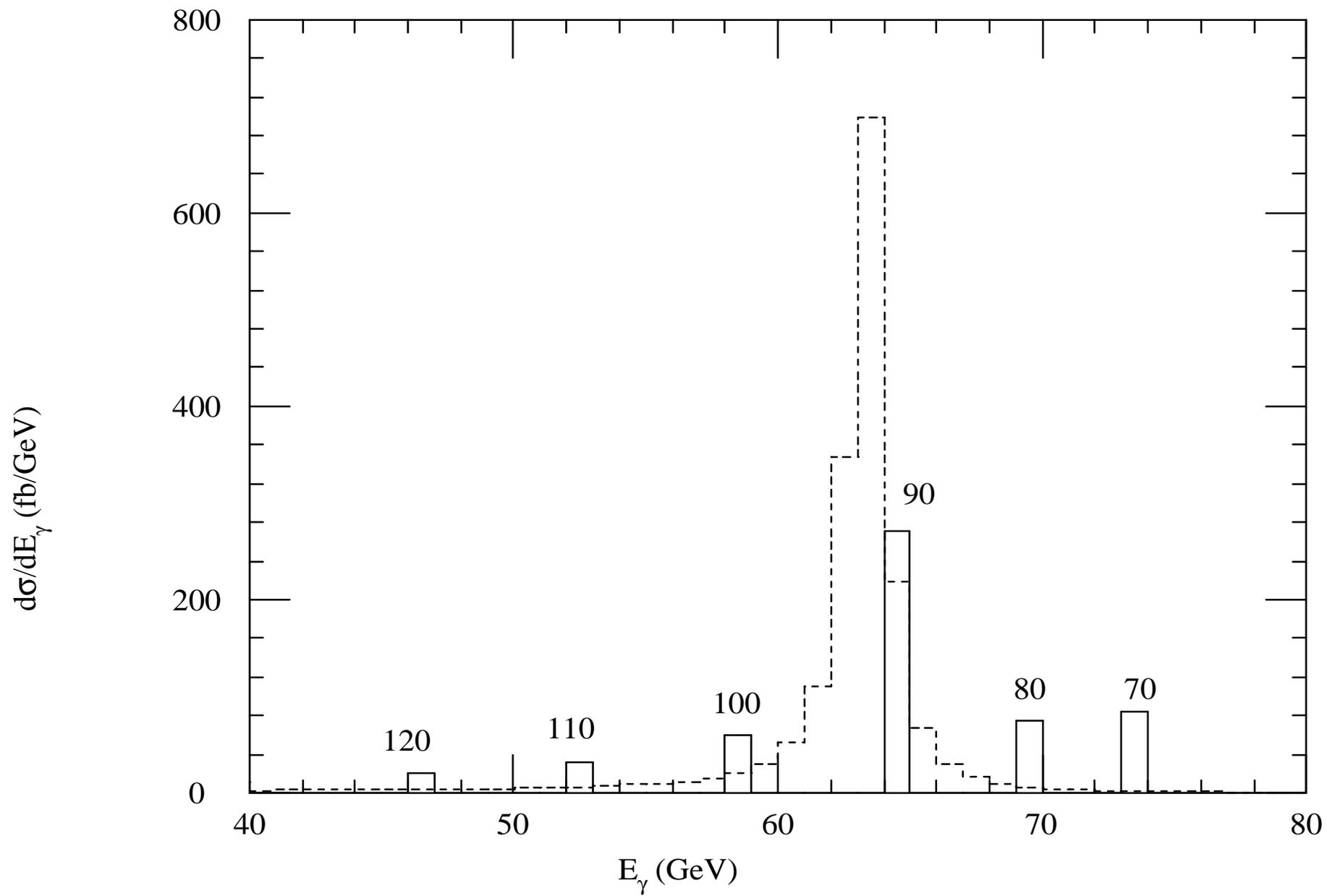
\begin{figure}
\protect
\caption{Photon energy distribution ($d\sigma/dE_\gamma$) of the
process $e^+ e^- \to b \bar{b} \gamma$, for the standard model
contribution (dashed histogram). We also show the Higgs boson
peaks (solid histogram) for different values of its mass, and
$g_{H \gamma \gamma} =  10^{-3}$ GeV$^{-1}$.}
\label{fig:2}
\end{figure}

Figure \ref{fig:2} shows our {\it typical} results for the photon
energy distribution presented as a $1$ GeV bin histogram. We have
taken  $g_{H \gamma \gamma} = 10^{-3}$ GeV$^{-1}$,
$g^{(1,2)}_{HZ\gamma} = 0$ and varied the Higgs mass between 70
and 120 GeV. We should point out that the general behavior of the
energy distribution remains the same when we consider the other
couplings, $g^{(1,2)}_{HZ\gamma}$, different from zero. We can
identify the $Z$--boson peak around $E_\gamma \simeq 64$ GeV and
also the various secondary peaks due to the Higgs boson at the
energies given by (\ref{ehiggs}). We can notice that the smaller
the Higgs mass, the larger is its effect in the $E_\gamma$
distribution.  Its detectability should rely on a careful
analyses of the tail (in the case where $M_H \neq M_Z$) of the SM
contribution to the photon energy spectrum in the process $e^+
e^- \to b \bar{b} \gamma$.

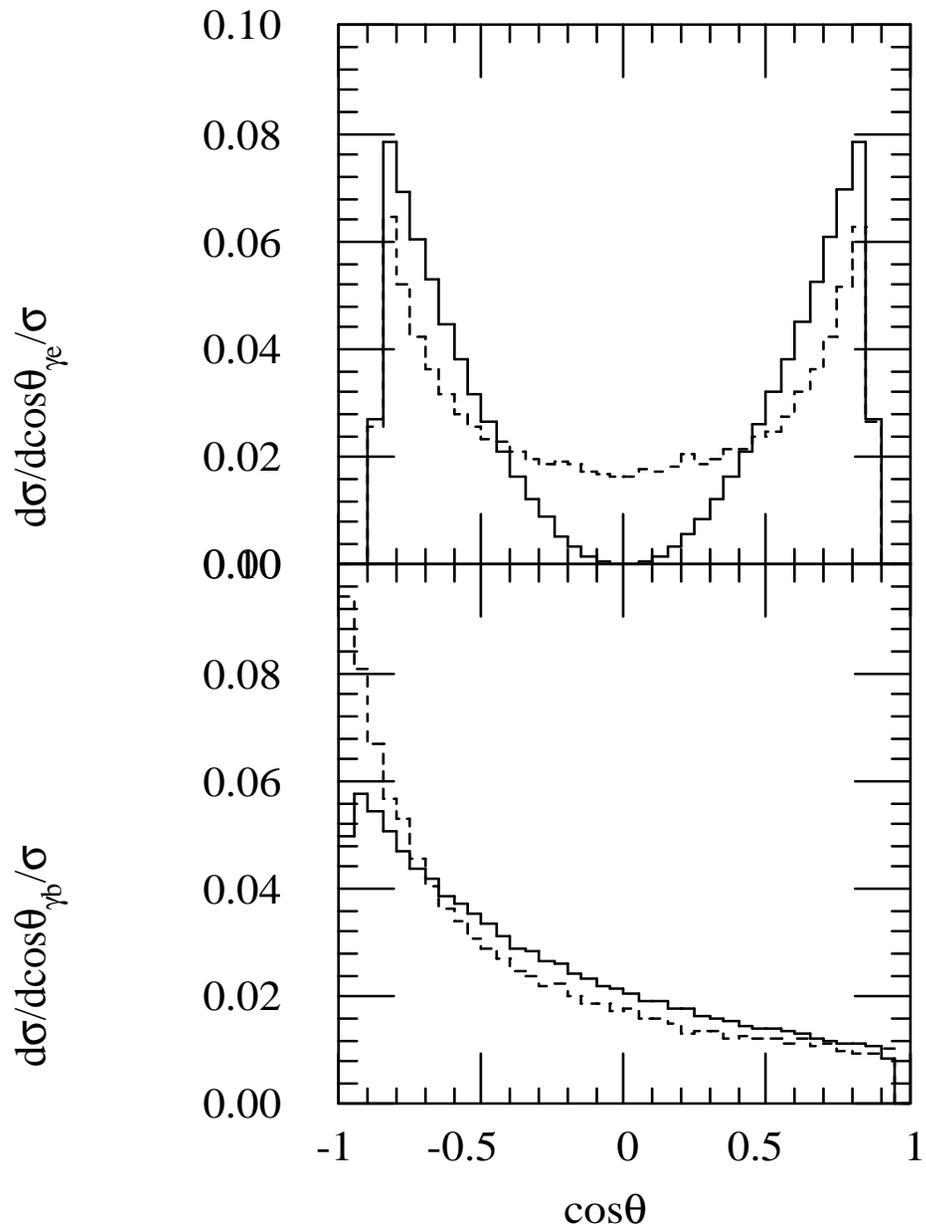
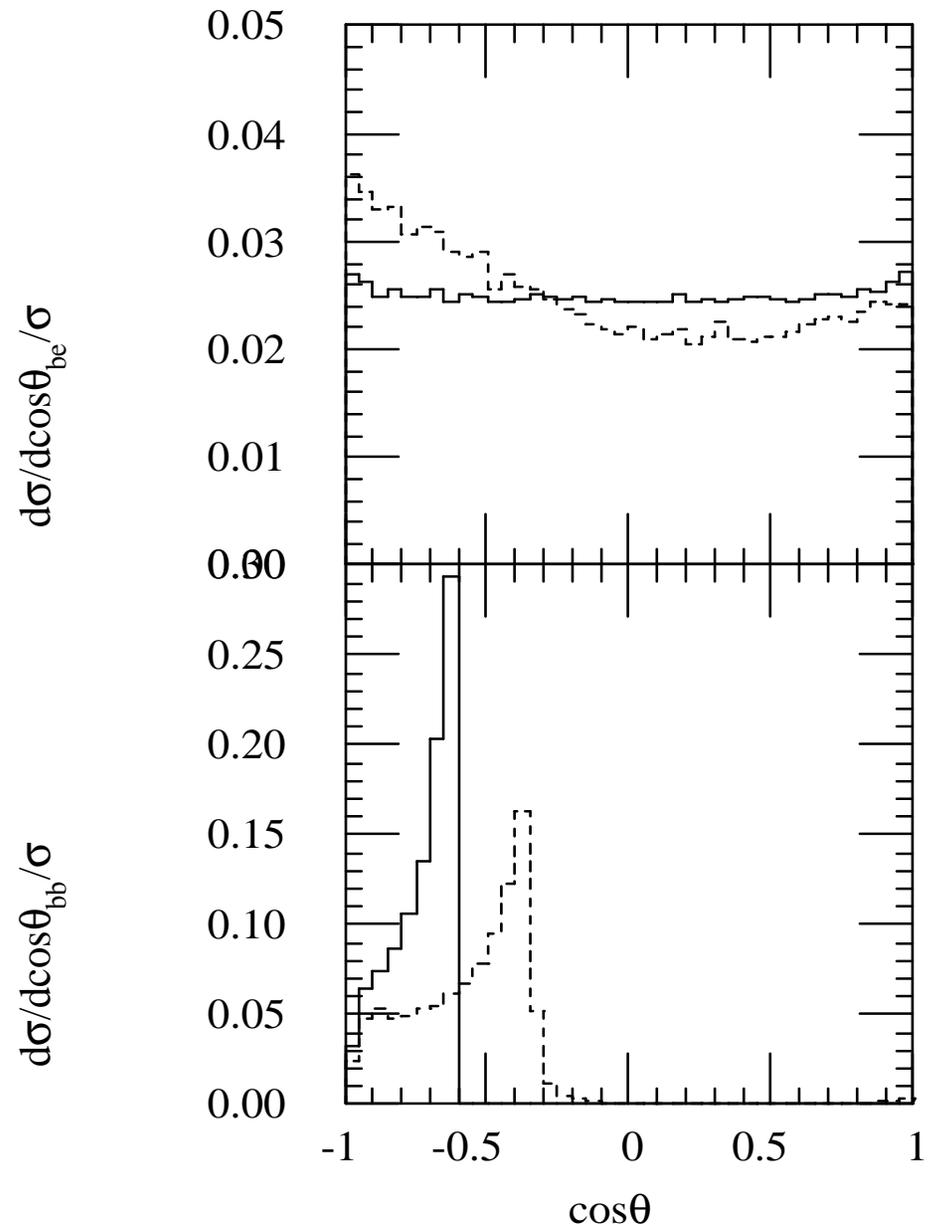
\begin{figure}
\protect
\caption{Normalized angular distribution $(1/\sigma)
d\sigma/d\cos\theta_i$, for $\theta_i = \theta_{\gamma e}$,
$\theta_{\gamma b}$, $\theta_{b e}$, and $\theta_{b \bar{b}}$.
The solid (dotted) lines represent the anomalous (standard model)
contributions, for $g_{H \gamma \gamma} =  10^{-3}$ GeV$^{-1}$.}
\label{fig:3}
\end{figure}

In Fig.\ \ref{fig:3}, for the sake of comparison between the
signal ($H$) and background ($Z$) behavior, we present separately
the normalized angular distribution for SM and anomalous
contributions, for $g_{H \gamma \gamma} = 10^{-3}$ GeV$^{-1}$,
and $M_H = 110$ GeV. We consider the angles between the electron
beam and the final particles ($\theta_{\gamma e}$, and $\theta_{b
e}$), and also the ones between the final particles
($\theta_{\gamma b}$, and $\theta_{b \bar{b}}$). We can see that
the signal has a very small contribution at $\theta_{\gamma e}
\sim \pi/2$, due to the scalar nature of the Higgs, whereas the
events from the background yields some events from transversal
$Z$'s in the central region. The $\theta_{b \bar{b}}$ angular
distribution shows that the produced quarks have a minimum angle
between themselves, which dependes on the mass and energy of
the parent particle, {\it i.e.\/} $\theta_{b \bar{b}} >
\theta^{\text{min}}_{Z(H)} = 2 \;  \arcsin (M_{Z(H)}/E_{Z(H)})$.
This variable could be used to further increase the signal over
background ratio. For instance, for $M_H > M_Z$, the cut
$\theta_{b \bar{b}} > \theta^{\text{min}}_{H}$ is able to get
ride of most of the $b \bar{b}$ events coming from the $Z$. In
fact, at the Z peak, the number of events is reduced by a factor
of 3 when this cut is implemented for a Higgs boson of 110 GeV.
On the other hand, for $M_H < M_Z$, a cut $\theta_{b \bar{b}} <
\theta^{\text{min}}_{Z}$ plays the same role. We should point out
that there is no difference among the distributions coming from
the three anomalous couplings (\ref{g}). Therefore, it will be
very difficult to make a distinction among these couplings based
only on these kinematical distributions \cite{epem}.

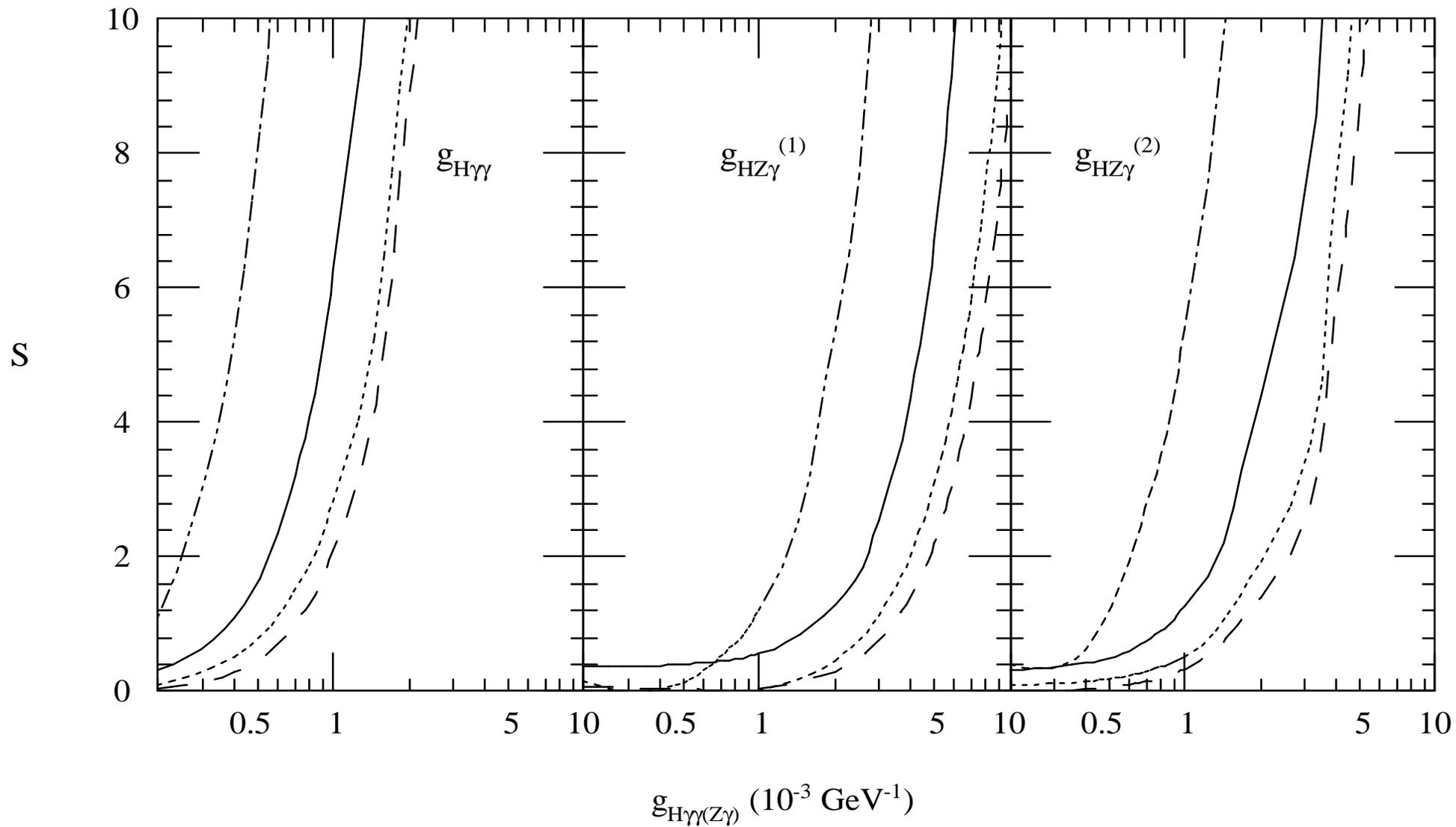
\begin{figure}
\protect
\caption{Significance of anomalous events as a function of the
coupling constants $g_{H \gamma \gamma}$, and $g^{(1,2)}_{H Z
\gamma}$, and different Higgs masses:  70 GeV (dot--dashed),  90
GeV (dashed), 110 GeV (solid), 130 GeV (dotted).}
\label{fig:4}
\end{figure}

In order to estimate the reach of LEP2 to disentangle the
anomalous Higgs boson couplings, we have evaluated the
significance ($S = \text{Signal}/\sqrt{\text{Background}}$) of
the signal based on the Higgs boson peaks in the $E_\gamma$
distribution, assuming a Poisson distribution for both signal and
background. We have scanned the parameter space for the three
anomalous couplings keeping only one non--zero coupling in each
run, for different values of the Higgs boson mass. We took the
coupling constants $g_{H \gamma \gamma}$, $g^{(1,2)}_{H Z
\gamma}$ in the range $10^{-4} - 10^{-2}$  GeV$^{-1}$
\cite{foot2}, and we assumed a $b$--tagging efficiency of 68\%
\cite{b:tag}. In Fig.\ \ref{fig:4} we present the significance
for each of these couplings, assuming four different values of
the Higgs boson mass $M_H = 70$, $90$, $110$, and $130$ GeV. We
should notice that for $M_H = 90$ GeV the signifance is reduced
due to the presence of the $Z$ boson peak. 

In Table \ref{tab:1}, we show the values of the coupling
constants $g_{H \gamma \gamma}$, $g^{(1,2)}_{H Z \gamma}$ that
corresponds to a $5 \; \sigma$ effect in the 1 GeV bin of the
$E_\gamma$ distribution around the Higgs peaks, for different
Higgs boson masses.  We also present the total number of signal
and background events in these bins. For $M_H = 90$ GeV, a large
numbers of events is needed due to the $Z$ boson peak. Since the
signal increases with the square of the anomalous couplings, for
some values of the coupling constants, we could expect to have a
reliable signal for the anomalous Higgs boson in less than one
year of LEP2 run.

In this study, we have not taken into account initial state
radiation, which would result in an energy degradation of the
original  $e^+ e^-$ beams, and we have not included a realistic
simulation of the  electromagnetic energy resolution. It is
important to notice that an increase in the $b$--tagging
efficiency, and a good resolution of the electromagnetic
calorimeter can help to select the $b \bar{b}$  events,
increasing the signal over background ratio and improving the
resolution of the Higgs boson peak in the photon energy
distribution.

In conclusion, searching for the anomalous Higgs at LEP2 provides
a complementary  way to the indirect precision measurements at
LEP1 in probing effective Lagrangians that are the low--energy
limit of physics beyond the SM. We have shown that the study of
the process  $e^+ e^- \to b \bar{b} \gamma$ can be a very
important tool in the search of these particles at LEP2. We found
that anomalous couplings  $g_{H \gamma \gamma}$, $g^{(1,2)}_{H Z
\gamma}  \sim 10^{-2}$ GeV$^{-1}$ are necessary for
identifying an anomalous Higgs of $150$ GeV. However, for a
lighter Higgs boson, couplings as small as $4 \times 10^{-4}$
GeV$^{-1}$ should suffice. 

\acknowledgments

We would like to thank O.\ J.\ P.\ \'Eboli for useful
discussions. This work was supported by Conselho Nacional de
Desenvolvimento Cient\'{\i}fico e Tecnol\'ogico (CNPq), and by
Coordena\c{c}\~ao de Aperfei\c{c}oamento de Pessoal de N\'{\i}vel
Superior (CAPES).


\widetext
\begin{table}
\begin{tabular}{||c||c||c||c||c||}
$M_H$ (GeV) & $|g_{H \gamma \gamma}|$ (GeV$^{-1}$) & 
$|g^{(1)}_{H Z \gamma}|$ (GeV$^{-1}$) & $|g^{(2)}_{H Z \gamma}|$ (GeV$^{-1}$) 
&  Signal / Background \\
\hline 
\hline 
70 & $3.89 \times 10^{-4}$ & $1.93 \times 10^{-3}$ & $9.62 \times 10^{-4}$ 
& 4.69/0.88 \\
\hline 
\hline
80 & $5.63 \times 10^{-4}$ & $2.66 \times 10^{-3}$ & $1.37 \times 10^{-3}$ 
& 7.32/2.14 \\
\hline 
\hline 
90 & $1.52 \times 10^{-3}$ & $7.47 \times 10^{-3}$ & $3.74 \times 10^{-3}$ 
& 43.09/74.26 \\
\hline 
\hline 
100 & $1.04 \times 10^{-3}$ & $4.98 \times 10^{-3}$ & $2.49 \times 10^{-3}$ 
& 13.61/7.41 \\
\hline 
\hline 
110 & $8.90 \times 10^{-4}$ & $4.30 \times 10^{-3}$ & $2.14 \times 10^{-3}$ 
& 7.16/2.05 \\
\hline 
\hline 
120 & $1.02 \times 10^{-3}$ & $4.91 \times 10^{-3}$ & $2.43 \times 10^{-3}$ 
& 5.81/1.35 \\
\hline 
\hline 
130 & $1.36 \times 10^{-3}$ & $6.44 \times 10^{-3}$ & $3.24 \times 10^{-3}$ 
& 4.77/0.91 \\
\hline 
\hline 
140 & $2.70 \times 10^{-3}$ & $1.09 \times 10^{-2}$ & $5.45 \times 10^{-3}$ 
& 4.72/0.89 \\
\hline 
\hline 
150 & $5.16 \times 10^{-3}$ & $2.65 \times 10^{-2}$ & $1.24 \times 10^{-2}$ 
& 4.92/0.97 \\
\end{tabular}

\caption{Values of the anomalous couplings $g_{H \gamma \gamma}$,
$g^{(1)}_{H Z \gamma}$, and  $g^{(2)}_{H Z \gamma}$ corresponding
to a significance of 5 $\sigma$, and the ratio of the total
number of signal and background events.}
\label{tab:1}
\end{table}


\begin{references}
\bibitem{foot2} For all the coefficients of (\ref{eff}) identical
to $f$, we have $|g_{H \gamma \gamma}| \simeq  6.03$ GeV $\times
f/\Lambda^2$ and $|g^{(2)}_{H Z \gamma}| \simeq 7.66$ GeV $\times
f/\Lambda^2$, and in this case, $|g_{i}| = 10^{-4}$ GeV$^{-1}$ is
equivalent to $f/\Lambda^2 \sim 16.6 (13.1)$ TeV$^{-2}$,
respectively.
\end{references}
\end{document}